\documentclass[10pt,twocolumn,letterpaper]{article}
\usepackage{spconf,amsmath,graphicx,amsfonts}
\usepackage{stfloats}
\usepackage{dcolumn}
\usepackage{bm}
\usepackage{cite}

\title{CNN FEATURES BASED UNSUPERVISED METRIC LEARNING FOR NEAR-DUPLICATE VIDEO RETRIEVAL}
\name{Hao Cheng, Ping Wang, Chun Qi}
\address{School of Electronic and Information Engineering \\
Xi'an Jiaotong University, Xi'an, China, 710049}
\begin{document}
%
\maketitle
\begin{abstract}
As important data carriers, the drastically increasing number of multimedia videos often brings many duplicate and near-duplicate videos in the top results of search. Near-duplicate video retrieval (NDVR) can cluster and filter out the redundant contents. In this paper, the proposed NDVR approach extracts the frame-level video representation based on convolutional neural network (CNN) features from fully-connected layer and aggregated intermediate convolutional layers. An unsupervised metric learning is used for similarity measurement and feature matching. An efficient re-ranking algorithm combined with k-nearest neighborhood fuses the retrieval results from two levels of features and further improves the retrieval performance. Extensive experiments on the widely used CC\_WEB\_VIDEO dataset shows that the proposed approach exhibits superior performance over the state-of-the-art.
\end{abstract}
\section{Introduction}
\label{sec:intro}
\begin{figure*}[t]
\centering
\includegraphics[width=18cm]{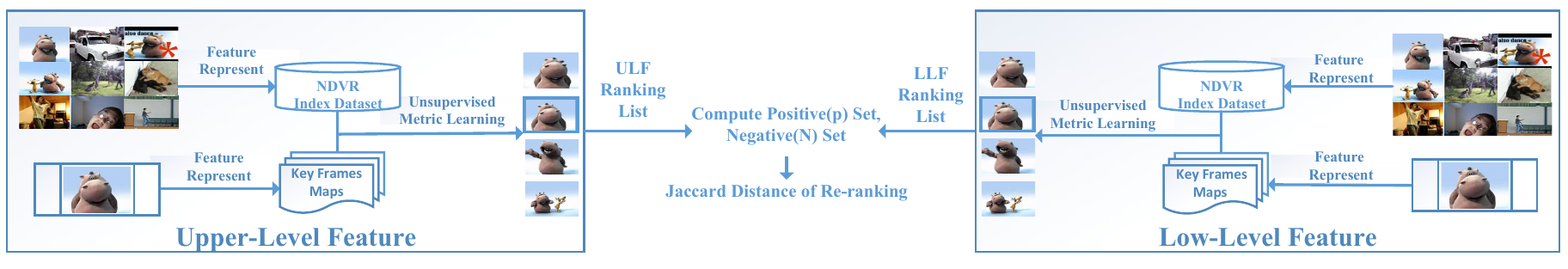}
\caption{\label{fig:wide}The outline of the proposed unsupervised near-duplicate video retrieval approach which includes three parts: feature extraction, unsupervised metric learning and re-ranking.}
\end{figure*}

In today's Internet era, multimedia videos serve as vital data carriers. With the increasing development of producing and spreading videos, a large number of duplicate and near-duplicate videos also inevitably appear. Wu et al.\cite{C1} searched on YouTube, Google Video and Yahoo! Video based on 24 keywords and obtained 12790 video results. They found that the average proportion rate of near-duplicate video reaches 27\%, and it is close to 93\% even for some category. This report shows that NDVR is an urgent problem to be solved.

Intuitively, two videos are defined to be near-duplicate if they are close to each other in content\cite{C2}. There are two critical steps for typical successful NDVR algorithms: feature representation for video and feature matching for similarity measurement of two individual videos\cite{C3}. Generally, video features can be divided into two levels: low-level features (LLF) including color, texture and SIFT features etc, and upper-level features (ULF) which have semantic information. Some global and local features of key frames are adopted in most NDVR algorithms, for example, Wu et al.\cite{C16} used the color histogram, Roover et al.\cite{C3} used color and marginal texture histograms, and Liu et al.\cite{C4} extracted the SIFT features of key frames. In order to obtain compact video signature, Song et al.\cite{C5} presented an approach called Multiple Feature Hashing (MFH), which mapped the video key frames into Hamming space, and tackled both the accuracy and scalability issues of NDVR. Hao et al.\cite{C6} proposed a novel stochastic multiview hashing (SMVH) algorithm to maximize a mixture of retrieval precision and recall scores. Recently, inspired by the outstanding performance of Convolutional Neural Network, Kordopatis et al.\cite{C7} leveraged the bag-of-word features of intermediate CNN outputs as final global video representation, and exhibited superior performance over various state-of-the-art approaches.

Most NDVR methods use single LLF. They often have large data volume for feature extraction. For feature matching, nearly all proposed methods perform exhaustive search based on simple Euclidean distance, Hamming distance\cite{C8} or others. These strategies do not fully utilize the interaction characteristic of pairwise videos. Some sophisticated algorithms are adopted to compensate this disadvantages. Chiu et al.\cite{C20} used dynamic programming to deal with the frame synchronization problem. Chou et al.\cite{C21} built a pattern-based indexing tree to do efficient retrieval. In this paper, we propose an efficient NDVR framework. In the step of feature extraction, CNN features based on pre-trained network models are used for video representation. To make use of CNN features, different levels of features including LLF and ULF are fused by a two-stage fusion strategy, which consists of numerical-based feature aggregation and semantic-based retrieval fusion. Further optimization of the performance is achieved via an efficient re-ranking algorithm. In respect of feature matching, we leverage an unsupervised metric learning framework to calculate the similarity of videos and kd-tree to store features and do fast retrieval. The proposed NDVR approach outperforms some well-known state-of-the-art, achieving a mean Average Precision (mAP) score of 0.9790.


The rest of this paper is organized as follows. Section 2 explains the details of the proposed NDVR approach. Section 3 reports the experimental results. Finally, section 4 concludes this paper.

\section{UNSUPERVISED NEAR-DUPLICATE VIDEO RETRIEVAL}
\label{sec:UNSUPERVISED NEAR-DUPLICATED VIDEORETRIEVAL STRUCTURE}
As shown in Fig. 1, the proposed unsupervised near-duplicate video retrieval approach will be described as feature representation, frame-specific unsupervised metric learning(FSUML) and re-ranking for semantic fusion.
\subsection{Feature Representation}
\label{sec:subhead}
The frame-level feature representation for a video consists of two steps as follows:

{\bf Key Frames Extraction:} According to \cite{C3,C16,C4,C5,C6,C7,C8,C20,C21}, video contents are accurately represented under a small amount of the data volume by a critical pre-processing named key frames selection. But commonly used random selection for key frames holds great randomness. In this paper, we extract key frames by two processes. Step 1):The difference of adjacent two frames is calculated based on the output vector of last CNN fully-connected layer. Then, the corresponding descending order of inter-frame distances for the whole video is made. $X=\{{x_i},i=1,2,...,n\}$ represents the frame sequences of a video with $n$ consecutive frames. $D=\{{D_i}=|{x_{i+1} - x_i}|,i=1,2,...,n-1\}$ represents the difference of adjacent two frames. Total number of key frames $m$ is determined according to a specific proportion, e.g. 2.5 frames per second. Corresponding difference of the $m$th position after descending order operation of $\{D_i\}$ is regarded as the threshold value which determines whether a frame is a key frame or not. Step 2):We further optimize the key frames set by comparing their temporal correlation and removing redundant frames when they are in the same second. The final key frames set is used to express video content.

{\bf Key Frames Representation:} Unlike the traditional Histogram\cite{C3}, SIFT\cite{C4} or Hash\cite{C5} features, a recent research work in\cite{C7} has put focus on intermediate layers of CNN network. In this paper, pre-trained CNN models are adopted to extract visual features from not only the intermediate convolutional layers but also the high fully-connected layer. The fully-connected layer has strong semantic information which has been successfully used for image classification. The output vector of the high fully-connected layer is considered as the upper-level video features. For the intermediate layers, there are multiple fusion strategies for features aggregation, e.g., Max pooling, Average pooling, Gram Matrix \cite{C9,C10}. Here, we use Maximum Activation of Convolutions (MAC)\cite{C11} to fuse the numerical-based intermediate layer features by operating the max pooling of CNN layer activations. A frame image is processed by a pre-trained CNN network with a total number of $L$ convolutional layers and generates a total of $L$ feature maps by forward propagation, denoted as $M$=$\{M^l\in\mathbb{R}^{{n_d^l} \times n_d^l \times c^l}$, $ l=1,2 \cdots L \}$, where $n_d^l \times n_d^l$ is the size of feature map for the $l$th convolutional layer, and $c^l$ is the total number of channels in each convolutional layer. MAC is to convert original feature maps to single feature descripor. The conversion process is formulated by $v^l(i) = max M^l(:, :, i), i=\{ 1,2, ... , c^l \}$, where every layer vector $v^l$ is a $c^l$-dimensional vector by max pooling on all channels of feature map $M^l$. Finally, low-level video features are composed of the MAC results of $L$ convolutional layer vectors through zero meaning averaging and $l2$ normalization.

\subsection{Frame-specific Unsupervised Metric Learning}
\label{ssec:subhead}
In order to reduce the storage requirements and processing time in querying, an efficient dimensionality reduction strategy named Kernel Principal Analysis (KPCA) is adopted before FSUML. KPCA can reduce nonlinear features to a low-dimensional level with sufficient separability (e.g. 4096 dimensions $\rightarrow$ 256 dimensions). We use Radial Basis Function(RBF) $K(x, x^{'}) = exp(- \frac{\| x-x'  \| {——^2}}{2\sigma^2})$ as project kernel.

Metric Learning can simultaneously integrate and conduct single or pairwise instances themselves to increase the natural separations among data samples. And unsupervised metric learning directly improves the similarity metric without introducing any extra distance notions and labels that are tagged in advance.

$q$ is the feature vector of one key frame of the query video, and $g$ is that of dataset video. In order to obtain the similarity between $q$ and $g$, metric learning is to learn a global Malhalanobis metric $M$ in following distance equation ${d(q,g)^2 = {(q-g)^T}M(q-g)}$.
In this paper, we use an unsupervised method called the simplified SSO\cite{C12} to obtain $M$. After KPCA, the dimensions of $q$ and $g$ are both 256. $q$=\{$q_i,i=1,...256$\}, $g$=\{$q_j,j=1,...256$\}, Similarity matrix $W$ with each entry $W(i,j) \in (0,1)$ represents the similarity between $q_i$ and $g_j$. The higher $W(i,j)$ is, the more similar $q_i$ is to $g_j$. $W$ is obtained by applying Gaussian kernel to a distance matrix as $W(i,j) = exp\{-d^2(i,j)/(k\sigma^2) \} $, where $d(i,j)$ denotes the distance between $q_i$ and $g_j$, and $k$ and $\sigma$ control the width of kernel. The Simplified SSO processing is as below:

\begin{itemize}
\item Firstly, computing the smoothing kernel $P$:

$P=D^{-1}W$, where $D$ is a diagonal matrix with $D(i,i) = \sum_{k=1}^{256}$ $W(i,k)$.
\item Secondly, performing smooth for $t$ steps:

$W_t = WP^t$, where $t$ is the number of times that execute this diffusion step. According to \cite{C13}, repeatedly multiplying by a matrix $P$ will bring the problem of Long-Term Dependencies. Consequently, we arrange $t$=1 and only smooth the similarity matrix $W$ once for preventing this problem.
\item Lastly, achieving self-normalization as:

$M^* = \Delta^{-1}W_t $, where $\Delta$ is a diagonal matrix with $\Delta(i,i)=\sum_{k=1}^{256}$ $W(i,k)$. This step guarantees the diagonal entries of smoothed similarity matrix are always 1.
\end{itemize}

After obtaining the metric matrix $M^*$, the final distance $d(q,g)_M^2$ via unsupervised metric learning can be rewrote as:
\begin{equation}
d(q,g)_M^2 =(q-g)^TM^*(q-g)
\end{equation}
The FSUML does not need any hard coding techniques, additional labels and other pre-processing algorithm.

\subsection{Semantic-based Fusion via Re-ranking }
\label{ssec:subsubhead}
Two aspects of fusion strategies are considered in this work, MAC above is used to fuse numeric intermediate convolutional layer features. Here, we will introduce a re-ranking algorithm to complete semantic-based multi-feature fusion.

Two levels of features from high fully-connected layer and intermediate convolutional layer are both extracted for the query video $q$ and the candidate video $g$. 
The unsupervised metric learning based on these features is used to measure the similarity between videos. According to the similarity, we look for two $k$-nearest neighborhood sets (KNNS), which are denoted as $N_k^{fc}(q)$ and $N_k^{conv}(q)$ based on two levels of features for the query video $q$.

{\bf Positive Set and Negative Set:}
The intersection set of two KNNS can be denoted as Positive Set (PS) \begin{equation}
N^{(PS)}_k(q) = N_k^{fc}(q) \cap N_k^{conv}(q)
\end{equation}
And call the union set as Negative Set (NS)
\begin{equation}
N^{(NS)}_k(q) = N_k^{fc}(q) \cup N_k^{conv}(q)
\end{equation}
The `positive' set means more correct near-duplicate video instances are covered and `negative' set contains more negative instance that is not near-duplicate. Actually, the principal task of re-rank is to further judge whether the query video $q$ is similar as candidate videos $g$ based on rank fusion.

Following \cite{C14}, we use Sparse Contextual Activation (SCA) to enhance retrieval performance. SCA is a highly efficient re-ranking algorithm algorithm, in which the neighborhood set $N_k(q)$ is converted to a vector representation $F_q=[F_{q,1}, F_{q,2},...,F_{q,N}]$ as
\begin{equation}
F_{q,g}=\left\{
\begin{aligned}
1 & \quad if\ g \in N_k(q) \\
0 & \quad otherwise
\end{aligned}
\right.
\end{equation}
$F_q$ has non-zero values only in the index where the neighbors of $q$ are located. It is a sparse vector. In order to perform extremely fast rank aggregation, for the query video $q$, two sparse contextual activations $F^{fc}_q$ and $F^{conv}_q$ based on two levels of features are generated, respectively. Then corresponding with positive set and negative set, two sparse contextual activations $F^{(PS)}_q$ and $F^{(NS)}_q$ are achieved.
\begin{equation}
F^{(PS)}_q =  MIN(F_q^{fc},F_q^{conv})
\end{equation}
\begin{equation}
F^{(NS)}_q =  MAX(F_q^{fc},F_q^{conv})
\end{equation}
The Jaccard distance of query video $q$ and candidate video $g$ for rank aggregation can be easily calculated by
\begin{equation}
d_J(q,g) = 1 - \frac{1}{2}\sum_{I=PS,NS}\frac{\sum_{i=1}^{N} min(F_{q,i}^{(I)},F_{g,i}^{(I)})}{\sum_{i=1}^{N} max(F_{q,i}^{(I)},F_{g,i}^{(I)})}
\end{equation}

Additionally, in order to perform fast retrieval, we use a modified kd-tree method called Randomly Projected kd-Trees\cite{C15} to obtain $k$-nearest neighborhood sets. This method has a good performance in high dimensional data space.

\section{Experiments}
\label{sec:Experiments}

{\bf CC\_WEB\_VIDEO}  is a well-known NDVR dataset consisting of 24 queries and 12,790 videos searched and downloaded from YouTube, Google Video and Yahoo! Video\cite{C16}. Two evaluation criterions including Precision-Recall (PR) curve and mean Average Precision (mAP) are adopted to evaluate the performance of the NDVR approaches.

\begin{equation}
\begin{split}
Recall= \frac {True\ Positive}{True\ Positive+False\ Negative}
\end{split}
\end{equation}
\begin{equation}
\begin{split}
Precision = \frac {True\ Positive}{True\ Positive+False\  Positive}
\end{split}
\end{equation}
where $Positive$ denotes the near-duplicate sample and $Negative$ means not near-duplicate sample.
\begin{equation}
mAP=\frac{1}{n}\sum_{j=1}^n\frac{1}{m}\sum_{i=0}^m \frac{i}{r_i}
\end{equation}
where $n$ is the number of query sets, $m$ is the number of relevant videos to the query video, and $r_i$ is the rank of the $i$-th retrieved relevant video.

\begin{table}[b]
\caption{mAP results with CNN structures and features}
\begin{center}
\begin{tabular}{|c||c|c|c|}
\hline
CNN Structure & LLF & ULF & RankF\\
\hline
AlexNet & 0.9734 & 0.9655 & 0.9780 \\
\hline
RCNN &  0.9737 & 0.9638 & 0.9758  \\
\hline
GoogLeNet &0.9782 & 0.9627  & {\bf 0.9790}\\
\hline
\end{tabular}
\end{center}
\end{table}

\begin{table*}[b]
\caption{Performance Comparison with the state-of-the-art}
\begin{center}
\setlength{\tabcolsep}{4mm}{
\begin{tabular}{|c||c|c|c|c|c|c||c|}
\hline
{\bf Method} & GF\cite{C16} & DP\cite{C20} & MFH\cite{C5} & PPT\cite{C21} & SMVH\cite{C6} & CNN-L\cite{C7} & CNN-UML\\
\hline
mAP& 0.892 & 0.900 & 0.928 & 0.958&0.971& 0.974 &{\bf 0.979}\\
\hline
\end{tabular}}
\end{center}
\end{table*}

{\bf Comparison with different CNN features:}
In this experiment, we show the results on CC\_WEB\_VIDEO dataset based on three pre-trained CNN architectures \{AlexNet\cite{C17}, RCNN\cite{C18}, GoogLeNet\cite{C19}\} in Table 1 and Fig. 2(a). ULF means features from the fully-connected layer, which is 4096-D from $fc7$ for AlexNet and RCNN and 1024-D from $pool5$ for GoogLeNet. LLF means features aggregation of the intermediate convolutional layers, which is 1376-D for AlexNet and RCNN and 5488-D for GoogLeNet. RankF denotes rank aggregation from two levels of features. It can be seen that both levels of features achieve good mAP results, and LLF has a little better performance compared with ULF for each CNN architecture. It illustrates two levels of features are both efficient for video representation. RankF derives the best results for all of three CNN architectures, and achieving the best performance for RankF of GoogLeNet. It indicates the rank fusion brings a little performance enhancement.


{\bf Comparison with existing NDVR approachs:}
Table 2 shows the performance of the proposed CNN based unsupervised metric learning (CNN-UML) approach and the six existing NDVR approaches which represent videos with different global or local features. Fig. 2(b) illustrates the PR curves of the compared approaches. The proposed CNN-UML outperforms all other methods, achieving the best mAP score 0.9790. Compared with the CNN-L approach which extracted bag-of-word features via codebook generation, CNN-UML exhibits better performance and does not need extra complex calculation.

\begin{figure}[t]

  \centering
  \centerline{\includegraphics[width=6.5cm]{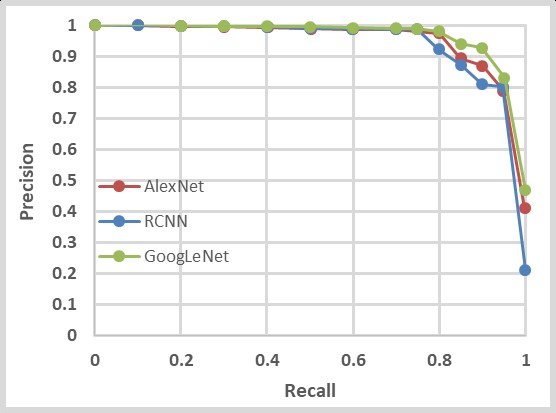}}
  \centerline{(a)}\medskip
\hfill
  \centering
  \centerline{\includegraphics[width=6.5cm]{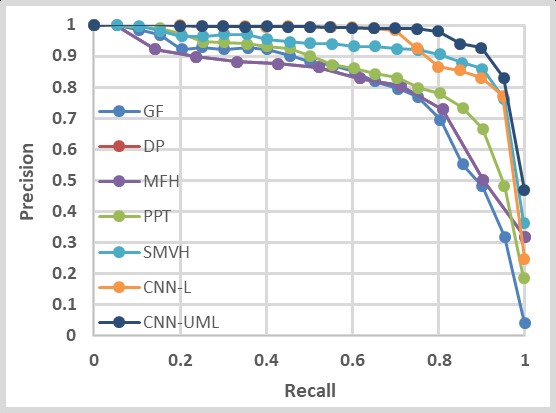}}
  \centerline{(b)}\medskip
%
\caption{Comparison of Precision-Recall Curves. (a) the best performances using three CNN structures (b) our best results and the state-of-the-arts}
\label{fig:res}
\end{figure}

\section{CONCLUSIONS}
\label{sec:CONCLUSIONS}
In this paper, we propose a near-duplicate video retrieval approach based on CNN features. Different levels of features from the fully-connected layer and the intermediate convolutional layers are used for video representation. Unsupervised metric learning and re-ranking algorithms are utilized to make the proposed approach more efficient. Experiments on a widely used CC\_WEB\_VIDEO dataset show that the proposed approach can effectively detect a large diversity of near-duplicate videos and filter out dissimilar ones. In the future, we plan to assess the performance on more challenging datasets. Furthermore, we will generalize the proposed approach to partial duplicate video retrieval task and also research an end-to-end supervised metric learning framework.


\bibliographystyle{IEEEbib}
\bibliography{strings} 

\begin{thebibliography}{10}

\bibitem{C1}
X.~Wu, C.W. Ngo, A.G. Hauptmann, and H.K. Tan,
\newblock ``Real-time near-duplicate elimination for web video search with
  content and context,''
\newblock {\em IEEE Trans. Multimedia}, vol. 11, no. 2, pp. 196--207, 2009.

\bibitem{C2}
J.~Liu, Z.~Huang, H~Cai, H.T. Shen, C.W. Ngo, and W.~Wang,
\newblock ``Near-duplicate video retrieval: Current research and future
  trends,''
\newblock {\em IEEE Multimedia}, vol. 45, no. 4, pp. 1--23, 2013.

\bibitem{C3}
C.D. Roover, C.D. Vleeschouwer, F.~Lefebvre, and B.~Macq,
\newblock ``Robust video hashing based on radial projections of key frames,''
\newblock {\em IEEE Trans. Signal Processing}, vol. 53, no. 10, pp. 4020--4037,
  2005.

\bibitem{C16}
X.~Wu, A.G. Hauptmann, and C.W. Ngo,
\newblock ``Practical elimination of near-duplicates from web video search,''
\newblock in {\em ACM International Conference on Multimedia (ACM MM)}. ACM,
  2007, pp. 218--227.

\bibitem{C4}
H.~Liu, H.~Lu, Z.~Wen, and X.~Xue,
\newblock ``Gradient ordinal signature and fixed-point embedding for efficient
  near-duplicate video detection,''
\newblock {\em IEEE Trans. Circuits \& Sysctems for Video Technology}, vol. 22,
  no. 4, pp. 555--566, 2012.

\bibitem{C5}
J.~Song, Y.~Yang, Z.~Huang, H.T. Shen, and J.~Luo,
\newblock ``Effective multiple feature hashing for large-scale near-duplicate
  video retrieval,''
\newblock {\em IEEE Trans. Multimedia}, vol. 15, no. 8, pp. 1997--2008, 2013.

\bibitem{C6}
Y.~Hao, T.~Mu, R.~Hong, M.~Wang, and et~al.,
\newblock ``Stochastic multiview hashing for large-scale near-duplicate video
  retrieval,''
\newblock {\em IEEE Trans. Multimedia}, vol. 19, no. 1, pp. 1--14, 2017.

\bibitem{C7}
G.~Kordopatis-Zilos, S.~Papadopoulos, I.~Patras, and Y.~Kompatsiaris,
\newblock ``Near-duplicate video retrieval by aggregating intermediate cnn
  layers,''
\newblock in {\em International Conference on Multimedia Modeling(MMM)}.
  Springer, 2017, pp. 251--263.

\bibitem{C8}
K.R. Kim, W.D. Jang, and C.S. Kim,
\newblock ``Frame-level matching of near duplicate video based on ternary frame
  descriptor and inerative refinement,''
\newblock in {\em IEEE International Conference on Image Processing(ICIP)}.
  IEEE, 2015, pp. 31--35.

\bibitem{C20}
C.Y. Chiu, C.S. Chen, and L.F. Chien,
\newblock ``A framework for handling spatiotemporal variations in video copy
  detection,''
\newblock {\em IEEE Trans. Circuits Syst. Video Technol}, vol. 18, no. 3, pp.
  412--417, 2008.

\bibitem{C21}
C.L. Chou, H.T. Chen, and S.Y. Lee,
\newblock ``Pattern-based near-duplicate video retrieval and localization on
  web-scale videos,''
\newblock {\em IEEE Trans. Multimedia}, vol. 17, no. 3, pp. 382--395, 2015.

\bibitem{C9}
J.~Johnson, A.~Alahi, and F.F. Li,
\newblock ``Perceptual losses for real-time style transfer and
  super-resolution,''
\newblock in {\em European Conference on Computer Vision(ECCV)}. Springer,
  2016, pp. 694--711.

\bibitem{C10}
L.A. Gatys, S.E. Alexander, and M.~Bethge,
\newblock ``Image style transfer using convolutional neural networks,''
\newblock in {\em IEEE Conference on Computer Vision and Pattern
  Recognition(CVPR)}. IEEE, 2016, pp. 1--16.

\bibitem{C11}
F.~Radenovic, G.~Tolias, and O.~Chum,
\newblock ``Cnn image retrieval learns from bow: Unsupervised fine-tuning with
  hard examples,''
\newblock in {\em European Conference on Computer Vision(ECCV)}. Springer,
  2016, pp. 3--20.

\bibitem{C12}
J.~Jiang, B.~Wang, and Z.~Tu,
\newblock ``Unsupervised metric learning by self-smoothing operator,''
\newblock in {\em IEEE International Conference on Computer Vision(ICCV)}.
  IEEE, 2011, pp. 794--801.

\bibitem{C13}
Ian. Goodfellow and et~al.,
\newblock ``Deep learning,''
\newblock {\em MIT Press}, http://www.deeplearningbook.org 2016.

\bibitem{C14}
S.~Bai and X.~Bai,
\newblock ``Sparse contextual activation for efficient visual re-ranking,''
\newblock {\em IEEE Trans. Image processing.}, vol. 25, no. 3, pp. 1056--1069,
  2016.

\bibitem{C15}
P.C. Wu, S.~HOI, D.D. NGUYEN, and Y.~He,
\newblock ``Randomly projected kd-trees with distance metric learning for image
  retrieval,''
\newblock in {\em International Conference on Multimedia Modeling(MMM)}.
  Springer, 2011, pp. 371--382.

\bibitem{C17}
A.~Krizhevsky, I.~Sutskever, and G.E. Hinton,
\newblock ``Imagenet classification with deep convolutional neural networks,''
\newblock in {\em International Conference on Neural Information Processing
  Systems}. Curran Associates Inc, 2012, pp. 1097--1105.

\bibitem{C18}
R.~Girshick, J.~Donahue, T.~Darrell, and J.~Malik,
\newblock ``Rich feature hierarchies for accurate object detection and semantic
  segmentation,''
\newblock in {\em IEEE Conference on Computer Vision and Pattern
  Recognition(CVPR)}. IEEE, 2014, pp. 580--587.

\bibitem{C19}
C.~Szegedy, W.~Liu, and Y.~Jia,
\newblock ``Going deeper with convolutions.,''
\newblock in {\em IEEE Conference on Computer Vision and Pattern
  Recognition(CVPR)}. IEEE, 2015, pp. 7--12.

\end{thebibliography}

\vspace{12pt}
\end{document}